\documentclass[10pt,journal,compsoc]{IEEEtran}
\ifCLASSOPTIONcompsoc
  \usepackage[nocompress]{cite}
\else
  \usepackage{cite}
\fi
\ifCLASSINFOpdf
\else
\fi
\usepackage{subfigure}
\usepackage{cite}
\usepackage{amsmath,amssymb,amsfonts}
\usepackage{graphicx}
\usepackage{textcomp}
\usepackage{xcolor}
\usepackage{fancyhdr}
\definecolor{link}{RGB}{120, 29, 125}
\usepackage{enumitem,kantlipsum}
\usepackage{algorithm}  
\usepackage{algorithmicx}  
\usepackage{algpseudocode}  
\usepackage{xcolor}
\usepackage{multirow}
\usepackage{pifont}
\usepackage{verbatim}
\usepackage{latexsym}

\usepackage[]{hyperref}

\usepackage[font={small, bf}, labelfont={small, bf}]{caption}

\begin{document}
%
\title{EnGN: A High-Throughput and Energy-Efficient Accelerator for Large Graph Neural Networks}
%
%
%
%

\author{Shengwen~Liang,
        Ying Wang, ~\IEEEmembership{Member,~IEEE,}
        Cheng Liu,
        Lei~He,
        Huawei Li, ~\IEEEmembership{Senior~Member,~IEEE,}
        and, Xiaowei Li, ~\IEEEmembership{Senior~Member,~IEEE}
\IEEEcompsocitemizethanks{\IEEEcompsocthanksitem The authors are with the State Key Laboratory of Computer Architecture, Institute of Computing Technology, Chinese Academy of Sciences, Beijing 100190, China, and also with the University of Chinese Academy of Sciences, Beijing 100190, China. E-mail:\{liangshengwen, wangying2009, liucheng, helei19g, lihuawei, lxw\}@ict.ac.cn} \protect\\
\thanks{Manuscript received April XX, 2020; revised August XX, 2020.}}

%
%

\markboth{IEEE TRANSACTIONS, VOL. X, NO. X, August~2020}%
{Shell \MakeLowercase{\textit{et al.}}: Bare Demo of IEEEtran.cls for Computer Society Journals}
%



\IEEEtitleabstractindextext{%
\begin{abstract}
Graph neural networks (GNNs) emerge as a powerful approach to process non-euclidean data structures and have been proved powerful in various application domains such as social networks and e-commerce. While such graph data maintained in real-world systems can be extremely large and sparse, thus employing GNNs to deal with them requires substantial computational and memory overhead, which induces considerable energy and resource cost on CPUs and GPUs. In this work, we present a specialized accelerator architecture, EnGN, to enable high-throughput and energy-efficient processing of large-scale GNNs. The proposed EnGN is designed to accelerate the three key stages of GNN propagation, which is abstracted as common computing patterns shared by typical GNNs. To support the key stages simultaneously, we propose the ring-edge-reduce(RER) dataflow that tames the poor locality of sparsely-and-randomly connected vertices, and the RER PE-array to practice RER dataflow. In addition, we utilize a graph tiling strategy to fit large graphs into EnGN and make good use of the hierarchical on-chip buffers through adaptive computation reordering and tile scheduling. Overall, EnGN achieves performance speedup by 1802.9X, 19.75X, and 2.97X and energy efficiency by 1326.35X, 304.43X, and 6.2X on average compared to CPU, GPU, and a state-of-the-art GCN accelerator HyGCN, respectively.
\end{abstract}

}

\maketitle

\IEEEdisplaynontitleabstractindextext

%
\IEEEpeerreviewmaketitle



\IEEEraisesectionheading{\section{Introduction}\label{sec:introduction}}
\IEEEPARstart{R}{ecently}, the success of deep learning methods in many fields has provoked a keen interest in generalizing neural network architectures to non-Euclidean data, such as manifolds and graphs. However, traditional deep neural networks, such as convolutional neural network (CNN)~\cite{ImageNet_Classification}, long short term memory (LSTM), are proposed to work for regular grid-like structures in Euclidean space, they are not trivially portable to non-Euclidean data domains like graphs. Therefore, graph neural networks (GNNs) are recently emerging as a powerful approach for graph processing and achieved unparalleled performance on many classic graph processing tasks, such as citation network~\cite{GCN}, social networks~\cite{FastGCN:Fast}, and knowledge graph~\cite{KnowledgeTransfer}. The success of graph neural networks propelled the deployment of GNNs to the real-world production system. For example, Alibaba's AliGraph~\cite{AliGraph} and Euler~\cite{Euler} platform leverage GNNs to analyze the e-commerce graph data of billion users and items. 

The prosperity of GNNs is enabling the development of emerging AI applications and systems that require high-throughput and low-latency processing capability. For instance, a recommendation system in Taobao~\cite{Euler} that leverages GNNs to mine billion-scale e-commerce data needs to perform real-time recommendations to millions of customers shopping at the same time. Therefore, to ease GNN model development and deployment, some high-performance GNN processing frameworks, such as Deep Graph Library (DGL)~\cite{DGL}, Pytorch Geometric (PyG)~\cite{PyG}, and Neugraph~\cite{NeuGraph:Parallel} have been developed because the existing deep learning frameworks and graph processing frameworks cannot fulfill large graph-based neural networks \cite{NeuGraph:Parallel}. However, the potential performance and energy efficiency of GNNs are still bounded by the hardware architectures assumed by these frameworks. The major drawbacks are attributed to three-fold factors. First, compared to DNNs with regular computing patterns, GNNs inherit both the irregular processing dataflow of graph analytic and the regular computing pattern of DNNs. This hybrid computing pattern that involves large amount of dynamic and irregular data accesses results in the inefficiency of the CPU and GPU. Second, a real-world graph can be extremely huge. For instance, the e-commerce graphs in Alibaba contain billions of nodes and hundreds of billion edges with rich attribute information. Some GNN software frameworks generally adopt a large number of compute nodes equipped with multiple CPUs or GPUs to deal with large-scale graphs, thus it results in high cost and energy overhead. For example, NeuGraph uses eight GPUs to handle a dataset with million vertices~\cite{NeuGraph:Parallel}. Third, the power-law distribution of the big real-world graph challenges the existing memory hierarchy and caching policy of CPUs and GPUs, for the sparsely distributed low-degree vertices in the graphs make it hard to reuse the graph data in general-purpose processors.

Intuitively, specialized hardware architecture is a promising option to improve the efficiency of GNN in together with GNN frameworks. However, previous graph processors and neural network accelerators are optimized to support either graph processing or neural networks, rather than both of them simultaneously. To address this problem, prior work proposed HyGCN to combine the graph processing and neural network processing in a specified hardware architecture. However, HyGCN mainly targets at graph convolution network (GCN) and utilizes a systolic array to perform neural network computation operation inside GCN, which is the target workload evaluated in their work, and it is not designed to run general GNN architectures like graph recurrent network, graph attention network, and etc. This is because the systolic array adopted by HyGCN is subject to low resource utilization when handling GRN with GRU or LSTM unit. In addition, the inherent nature of the real-world large-scale graph adopted by the GNN model, such as power-law distribution, variable feature length of vertex, significantly impacts the performance of GNN model and it leaves a large potential space to optimize the data locality, on-chip memory hierarchy, task partitioning, and scheduling. Nevertheless, HyGCN is more suitable for moderate-scale graphs and does not consider such inherent features of the graph, which dramatically limits the achieved performance and energy efficiency.    

Therefore, in order to solve the aforementioned issues and accelerate practical GNN-based applications that process real-world large-scale graphs, we propose EnGN, a high-throughput and energy-efficient edge-centric accelerator for large graph neural network processing. However, design such an accelerator is a non-trivial task and has to resolve the obstacles that exist in the real-world GNN algorithms: (1) How to tailor a unified architecture that efficiently supports the diverse GNN models and flows not limited to GCNs. It is observed that the dataflow and the dimension of the working-set, e.g., the vertex, dynamically changes in wide ranges during the propagation of different GNN layers, requiring a reconfigurable architecture and interconnects to avoid hardware and memory bandwidth under-utility. (2) large graphs containing millions of vertices pose a significant challenge to the design of energy-efficient and compact GNN accelerators with limited on-chip memory space. Particularly, when massive graphs with million vertices are partitioned into sparsely-connected sub-graphs, there will be intensive random and irregular off-chip memory accesses induced, which leads to poor locality that are hard to harness in the aggregate and update stage. And (3) the power-law distribution~\cite{Powerlaw} creates high-degree but imbalanced connection sparsity in large real-world graphs. Accelerator must be able to deal with the imbalanced sparsity distribution, which leads to processing elements under-utility, poor locality, and redundant memory access issues in hardware.

To cope with issues, first, by observing state-of-the-art GNN processing frameworks such as DGL and PyG, we generalize the architecture of typical GNN algorithms into three key stages: the vertex feature extraction stage, the feature aggregate stage, and the graph update stage. In response to the three key stages abstracted from general GNN frameworks, we support the corresponding computing patterns in EnGN, so that it is a general GNN processor and able to support most of the GNN architectures such as GCN, GRN, and etc. In EnGN, a ring-edge-reduce (RER) dataflow and the accompanied hardware architecture of RER processing elements (PEs) arrays are designed to simultaneously conduct the stages of vertex property feature extraction, aggregate, and vertex update on GNNs. It is known that aggregating the property and updating the vertices distributed in the large but sparse graphs will lead to poor hardware resources and memory bandwidth utility due to poor data locality of vertices and edges. However, the proposed RER PEs connected into a ring topology leverages the RER dataflow to make vertex property flow between rows of PEs and performs efficient update operations without randomly accessing the vertices and edges from the memory.

\begin{table*}[!t]
\centering
\footnotesize
\setlength{\abovecaptionskip}{5pt}
\caption{GNN algorithms on EnGN processing model.}
\label{table:GNN algorithms on GNN framework}
\begin{tabular}{ccccc}
\hline
Algorithms             & Feature extraction     & Aggregate      & Update        \\ \hline
GCN                    & $h^{l}_{u} \ast V^{-1/2}_{degree}               $   & $V^{l}_{temp} = acculumate(Res)   $          &  $ReLu(W^{l}V^{l}_{temp})                    $        \\ 
GS-Pool         & $ReLu(W^{l}_{pool}V^{l}_{u}+b)                   $   & $V^{l}_{temp} = max(Res)          $          &  $ReLu(W^{l}concat(V^{l}_{temp}, h^{l}_{v}))  $        \\ 
R-GCN                  & $h^{l}_{r,u} \ast V^{-1/2}_{degree}             $   & $V^{l}_{r,temp} = acculumate(Res) $          &  $ReLu(\sum_{r\in R}W^{l}_{r}V^{l}_{r,temp}) $        \\ 
Gated-GCN     & $Sigmoid\ (W_H^lh_v^l+W_C^lh_u^l)\ \odot h_u^l  $   & $V_{temp}^l=acculumate(Res)       $          &  $ReLu\ (W^l\ V_{temp}^l)                    $        \\ 
GRN                    & $h_u^l                                          $   & $V_{temp}^l=acculumate(Res)       $          &  $GRU(h_{v}^{\left(l\right)},\ W^l\ V_{temp}^l)  $        \\ \hline
\end{tabular}
\vspace{-15pt}
\end{table*}

Second, for the feature extraction stage, EnGN constructs a graph property aware dataflow (GPA) that decouples the vertex property and the hardware structure, which makes the GNN mapping to the RER array independent of the vertex dimension. Meanwhile, due to the change of vertex-property dimension after the aggregate stage, how to schedule the three stages in GNN layers makes a significant impact on the total computation cost. Thus, GPA can dynamically reorder the graph processing stages to reduce the total computation cost of the GNN model on the accelerator. 

Third, considering the footprint of large-scale graphs, EnGN adopts a graph tiling strategy to process the partitioned sub-graphs with high degree of data reusability. Graph tiling aims to partition a large-scale graph into sub-graphs that fit the on-chip memory and maximize the locality. The tiles are strategically scheduled in EnGN to select either row-oriented or column-oriented processing dataflow to maximally reuse vertices between tiles and reduce the overhead caused by the off-chip memory access.

Finally, due to the power-law distribution and sparsity characteristics of the real-world graphs, the accessing frequency to different vertices may vary in a large scale. For example, on Cora citation graph~\cite{Cora}, the access frequency of a high-degree vertex is 100x times than that of a low-degree vertex, which causes access imbalance issue. Thus, EnGN comprises a three-level on-chip memory hierarchy, and the L2 memory is a degree-aware vertex cache (DAVC) to locally cache the high-radix vertices that are densely connected to other vertices in graphs. DAVC reduces considerable memory access cost. In summary, our main contributions are the following:
\begin{enumerate}
	\item A compact but high-throughput accelerator is designed for large graph neural network, which is implemented based on the edge-centric paradigm and supports various large scale GNNs.
	\item We proposed a graph property aware and ring-edge-reduce (RER) dataflow to enable the EnGN to handle a vertex with arbitrary dimension property and high throughput update operations. The on-chip memory hierarchy is designed to be aware of the non-uniform distribution of high-radix and low-radix graph vertexes and employ a specialized memory space management to enhance data locality on the chip.
	\item We implement the EnGN accelerator in 14nm process and make comprehensive evaluations and compare the performance, power, energy of EnGN to CPU, GPU, and HyGCN baselines. Experimental results show that, compared to CPU and GPU, EnGN achieves on average 1802.9X speedup with 1326.35X energy reduction and 19.75X speedup with 304.43X energy reduction, respectively. The speedup and energy efficiency of EnGN is shown to be 2.97X and 6.2X higher than HyGCN, which is a contemporary work of EnGN on GNN accelerator.  
\end{enumerate}

\section{General GNN processing model}
\label{sec:section_2}
\subsection{Graph neural networks}
Unlike convolutional neural networks that mainly deal with Euclidean data like images and videos~\cite{NeuGraph:Parallel}, graph neural networks (GNNs) generalize the conventional neural networks to operate directly on non-Euclidean data especially graph data such as social networks and chemical molecules. It has been proven to be supremely successful on tasks like node classification, graph classification, and link prediction. Motivated by the success of GNNs, various GNN architectures have been proposed recently \cite{GraphNN_survey,GNN_survey}.

\textbf{Graph convolution network (GCN)} generalizes the convolution operation from regular image data to non-structural graph data. It can be used for node classification~\cite{GCN} and chemistry molecules architecture analysis~\cite{chemistry_GNN}. A typical GCN~\cite{GCN} is presented and formulated in~\autoref{equ:GCN}:
\begin{equation}\label{equ:GCN}
\small
\setlength{\abovedisplayskip}{3pt}
\setlength{\belowdisplayskip}{3pt}
h^{l+1} = ReLu(\tilde{D}^{-1/2}\tilde{A}\tilde{D}^{-1/2}h^{l}W^{l}), h^{0} = X
\end{equation}
Note that $A$ is the adjacency matrix of the graph, $W^{(l)}$ is the weight matrix at layer $l$, 
$\tilde{D}_{ii} = \sum_{j}\tilde{A}_{ij}$ is essentially the output of the normalized graph 
Laplacian~\cite{GCN} over $A$ where $I_N$ is the identity matrix and $\tilde{A}=A + I_{N}$.

\textbf{GraphSage-Pool (GS-Pool)} is proposed in~\cite{Reddit} and used for citation network analysis and protein-protein interaction task. Unlike the GCN models, it leverages the averaging function as an aggregation operator and has the source
vertex property ($h^{l}_{v}$) involved when updating output in next iteration. The expression of GS-Pool is defined in~\autoref{equ:graphsage-pool}.
\begin{equation}\label{equ:graphsage-pool}
\small
\setlength{\abovedisplayskip}{3pt}
\setlength{\belowdisplayskip}{3pt}
h_{v}^{l+1} = ReLu(W^{l}concat(ReLu(W^{l}_{pool}V^{l}_{u}+b)), h^{l}_{v})
\end{equation}
where $concat(\cdot)$ acts as the function that concatenates a vertex's 
property with the aggregated property of its neighbor vertices and $V^{l}_{u}$ is the source vertex property at layer $l$.

\textbf{Relational graph convolutional network  (R-GCN)} is an extension of GCN and used to handle graphs with different edge types. For instance, the edges can be used to represent different relations and have distinct weights definition of $W^{l}_{r}$ ~\cite{RGCN}. Similar to GCN, hidden representation 
of entities in the $(l+1)^{th}$ layer in R-GCN can be formulated in~\autoref{equ:RGCN}:
\begin{equation}\label{equ:RGCN}
\small
\setlength{\abovedisplayskip}{3pt}
\setlength{\belowdisplayskip}{3pt}
h^{l+1}_{i} = \sigma{(W^{l}_{0}h^{l}_{i} + \sum_{r\in R}\sum_{j\in N^{r}_{i}}\frac{1}{c_{i,r}}W^{l}_{r}h^{l}_{j})}
\end{equation}
where $N^{r}_{i}$ denotes the set of neighbor indices of node $i$ under relation $r\in R$ 
and $c_{i,r}$ is a normalization constant. $c_{i,r} = |N^{r}_{i}|$ is used in prior 
entity classification work~\cite{RGCN}.

\textbf{Gated graph convolution network (Gated-GCN)} is proposed in~\cite{Gated_Conv} and
utilized for community detection. It borrows the idea from gate recurrent neural 
networks and constructs a propagation function that receives and processes the property of source vertex and destination vertex simultaneously. The propagation function is depicted in~\autoref{equ:gate_gcn}.
\begin{equation}\label{equ:gate_gcn}
\small
\setlength{\abovedisplayskip}{3pt}
\setlength{\belowdisplayskip}{3pt}
\begin{split}
	h_{v}^{\left(l+1\right)} &=\ Relu\ (W^l(\sum_{u\in N(v)}{\eta_{uv}\odot h_u^l}) \\
	\eta_{uv} &=\ Sigmoid\ (W_H^lh_v^l+W_C^lh_u^l)
\end{split}
\end{equation}
where $\odot$ refers to element-wise multiplication, $ReLu(\cdot)$ and $sigmoid(\cdot)$ are typical 
nonlinear activation functions that have been widely adopted in CNNs~\cite{ImageNet_Classification}. 

\textbf{Graph Recurrent network (GRN)} is similar to the recurrent neural network (RNN), but aims to learn vertex representations~\cite{RNNnrework}. GRN is mostly used in NLP tasks, traffic forecasting, and etc. For example,~\cite{GRU_Graph} integrates typical RNN units (Gated recurrent unit) into the propagation function as formulated in~\autoref{equ:grn} to perform graph learning tasks.
\begin{equation}\label{equ:grn}
\small
\setlength{\abovedisplayskip}{3pt}
\setlength{\belowdisplayskip}{3pt}
    h_{v}^{\left(l+1\right)}=\ GRU(h_{v}^{\left(l\right)},\ \sum_{u\in N(v)}{W^{l}h_u^l})
\end{equation}
Although GNN algorithms are different in terms of architecture and target applications, we notice that they share common computing patterns. 1) GNNs initially condense vertex property of source vertex with learned parameters to obtain more compact feature representations. 2) Afterwards, GNNs usually gather neighbor properties to embed the information of graph topology to the extracted features. and 3) GNNs usually leverage learned parameters to condense the output features obtained in the aggregate stage making GNN capable to learn and perform more complex tasks. GNN accelerators must be able to support the computation abstractions concluded above, in order to support different GNN architectures efficiently.
\setcounter{algorithm}{0}
\begin{algorithm}[h]
\scriptsize
   \begin{algorithmic}[1]
	\caption{{EnGN processing model}}
	\label{alg:EnGN-model}
	\Require Graph $G = (V,E)$, Vertex property $Prop$ and $Tmp_{prop}$, layer $l$, learned parameter $W_{feature}, W_{update}$
    \Ensure Vertex Property $Result$
    \For{$l \leftarrow 1 $  to  $l_{max}$}
		\For{each edge $e \in Edge$}
			\State $tmp \leftarrow \textcolor{red!70!black}{\texttt{Feature\ extraction}}(Prop[e.src], Prop[e.dst], W_{feat.})$ 
			\State $Tmp_{prop}[e.dst] \leftarrow \textcolor{red!70!black}{\texttt{Aggregate}}(Tmp_{prop}[e.dst],tmp)$
		\EndFor
		\For{each edge $e \in Edge$} 
			\State $Prop[e.dst] \leftarrow \textcolor{red!70!black}{\texttt{Update}}(Prop[e.dst], Tmp_{prop}[e.dst], W_{update})$
		\EndFor
	\EndFor
	    \State $Result \leftarrow Prop$
   \end{algorithmic}
\end{algorithm}
\vspace{-20pt}
\subsection{EnGN processing model}
\label{sec:processing model}
According to the goal of the key stages in a typical GNN, the common computing patterns can be 
abstracted as \textit{feature 
extraction}, \textit{aggregate}, and \textit{update}. The \textit{feature extraction} stage condenses the property 
of each vertex in the graph using a neural network. The \textit{aggregate} stage embeds the graph topology into local vertex property by accumulating each vertex's  neighbor properties generated in the feature extraction. The choices of aggregate functions include various 
arithmetic operations such as max, min, and add to produce unified output features. At the end of propagation iteration, the \textit{update} stage leverages learned parameters to further condense the output features obtained in the aggregate stage, then applied a non-linear activation function or GRU/LSTM function to each vertex of the graph before output. Note that when the aggregate stage includes only linear operation, it can be scheduled before or after the feature extraction stage. It also provides an opportunity for EnGN to dynamically adjust the stages of matrix operations to optimize EnGN performance, which will be introduced in~\autoref{sec:optimization}.
On top of the abstraction, we propose a unified EnGN processing model that can cover general GNN models 
using the common computing functions as shown in \autoref{alg:EnGN-model}. Suppose the graph is 
represented as $G(V, E)$ where $V$ and $E$ represent the set of vertices and edges in the graph respectively.
$Property$ is the set of vertex property of the graph. By default, the input graph is stored as a coordinate list (COO). 
Each edge in the graph is a tuple ($src$, $dst$, $val$), where $val$ usually stands for the edge property and it depends on graph definition. 
The EnGN execution flow follows the neighborhood aggregation strategy, which iteratively updates the 
representation of vertices by aggregating representations of their neighbors. Since all the vertices 
in the graph will be processed in each iteration for GNN algorithms, EnGN is presented as an edge-centric processing model 
to ensure more efficient memory accesses~\cite{Edge_centric}. 

For each edge, both the source vertex property and the destination vertex property are condensed with 
$W_{feature}$ using $feature\_extraction(\cdot)$ to obtain a temporary property $tmp$.
Then $tmp$ is added to the destination property using $aggregate(\cdot)$ function.
Since there may be multiple edges that are incident to the same destination vertices, $aggregate(\cdot)$ 
is essentially a reduce function. When all the destination vertices are reduced, an activation function or the user-defined operator with learnable weights $W_{update}$ are used to filter the output using $update(\cdot)$ function.


\begin{figure}[!t]
\centering
\setlength{\abovecaptionskip}{0pt}
\includegraphics[width=0.9\columnwidth]{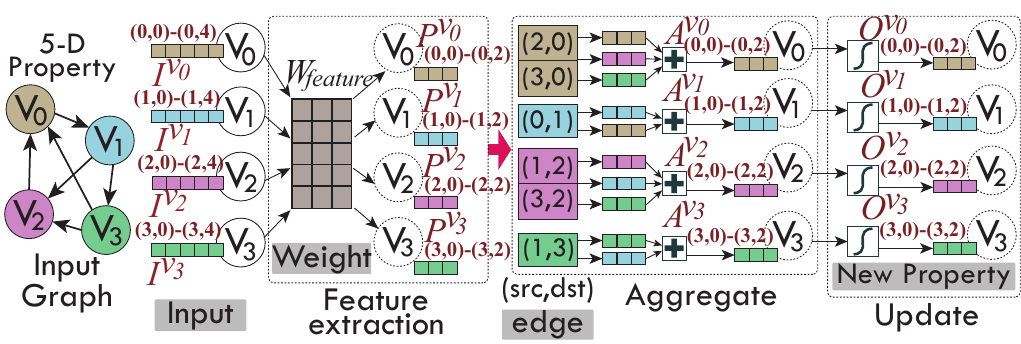}
\caption{GCN on EnGN processing model.}
\label{fig:EnGN processing model}
\vspace{-15pt}
\end{figure}

To help understand the EnGN execution model, we present a vivid example of GCN~\cite{GCN} processed by the EnGN architecture
as shown in~\autoref{fig:EnGN processing model}. Suppose an input graph has four vertices and 
each vertex has a 5-dimension property. The input property of the vertices are denoted 
as $I^{v_0}$, $I^{v_{1}}$, $I^{v_{2}}$, $I^{v_{3}}$.
In $feature\_extraction(\cdot)$ function, the feature extraction function takes both the vertex 
property i.e. $I^{v_0}$, $I^{v_{1}}$, $I^{v_{2}}$, $I^{v_{3}}$ and associated 
weight matrix $W_{feature}$ as input. Then it has the weight matrix multiplied with 
the high-dimension input vertex property to generate low-dimension temp features. 
Note that the size of the weight matrix is associated with both the input property dimension 
and output temp feature dimension. In this example, the size of the weight matrix is $5\times3$. 
With the feature extraction function, the input vertex properties are converted to 3-dimension 
temp features donated as $P^{v_0}$, $P^{v_{1}}$, $P^{v_{2}}$, $P^{v_{3}}$.
In $aggregate(\cdot)$ function, it receives the results of $feature\_{extraction}$ function 
and aggregates the property of each vertex's incoming neighbors. As shown in~\autoref{fig:EnGN processing model}, 
the temp properties of vertex 2 and 3 i.e. $P^{v_{2}}$, $P^{v_{3}}$ are added to temp property of vertex 0 as vertex 2 and 3 are incoming neighbors of vertex 0 $P^{v_{0}}$ according to the graph topology. When the aggregation stage is done, $update(\cdot)$ starts. It has the vertex features i.e. $A^{v_{0}}$, $A^{v_{1}}$, $A^{v_{2}}$, $A^{v_{3}}$ filtered using an activation function. The filtered output properties denoted as $O^{v_{0}}$, $O^{v_{1}}$, $O^{v_{2}}$, 
$O^{v_{3}}$ become the input to the next iteration.

Similar to the GCNs, we also have the rest of the typical GNN algorithms mentioned in~\autoref{sec:section_2} mapped 
to the EnGN processing model.~\autoref{table:GNN algorithms on GNN framework} summarizes the resulted EnGN processing 
functions.

\section{Motivation}

\begin{figure}[!t]
\centering
\setlength{\abovecaptionskip}{0pt}
\includegraphics[width=0.9\columnwidth]{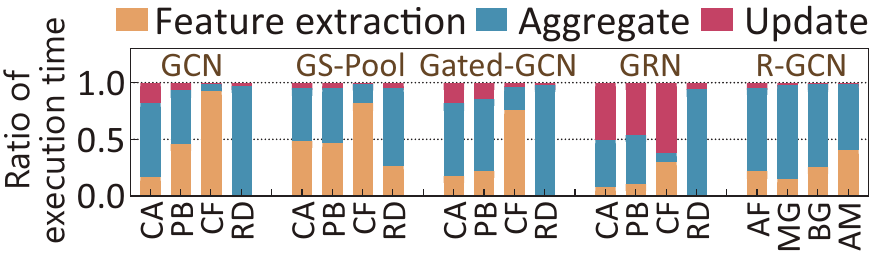}
\caption{Execution time breakdown of GNN models.}
\label{fig:execution time}
\vspace{-15pt}
\end{figure}

\subsection{Workload characterization}
To gain insight into the computing characteristics of GNN processing models, we leverage a state-of-the-art GNN software framework, DGL, to analyze the five aforementioned GNN models on Intel Xeon CPU. \autoref{fig:execution time} shows the execution time breakdown of GCN, GS-POOL, Gated-GCN, GRN, and R-GCN on the datasets that are selected from~\autoref{table:dataset}. Note that GCN, GS-POOL, Gated-GCN, and GRN executed on datasets of CA, PB, CF, and RD while R-GCN is mainly used in the knowledge graph and it works on open datasets including AF, MG, KG, and AM. In general, it can be observed that the three processing stages including feature extraction, aggregate, and update take up a distinct proportion of the execution time on different datasets. Thereby, all the processing stages must be taken into consideration for general GNN acceleration which remains a great design challenge. On the other hand, we observe that the aggregate stage that requires computing and traverse of the graph data involves considerable irregular memory accesses and consumes a large portion of the total execution time on datasets of CA, PB, and RD for algorithms of GCN, GS-Pool, and Gated-GCN. Particularly, the aggregate stage of R-GCN on all the datasets turns out to be the most time-consuming stage. To further investigate the reasons for the processing inefficiencies of the aggregate processing stage, we analyze the statistics of the CPU processing system executing GNNs as listed in \autoref{table:exec_pattern}. The results reveal that the aggregate stage has the lowest instructions per cycle (IPC) due to the much higher cache miss rate and memory bandwidth requirements, which are mostly incurred by the intensive irregular memory accesses. According to the I/O to computing ratio metric i.e. memory accesses per operation in the table, we also confirm that the aggregate stage involves intensive memory accesses per operation. In a nutshell, the aggregate stage closely relevant to the irregular graph is the most critical part of GNN processing in most cases. It is an IO-bound task and must be optimized sufficiently for high-performance GNN processing.

\begin{table}[!t]
    \scriptsize
    \centering
    \setlength\tabcolsep{4pt}
    \setlength{\abovecaptionskip}{0pt}
    \caption{Execution pattern of GCN on Cora dataset.}
    \label{table:exec_pattern}
\begin{tabular}{cccc}
\hline
\textbf{}                                                                   & \textbf{Feature extraction} & \textbf{Aggregate} & \textbf{Update} \\ \hline
IPC                                                                         & 1.73                        & \textbf{0.77}      & 1.01            \\
L3 cache miss ratio                                                         & 56.60                       & \textbf{82.62}     & 46.47           \\
\begin{tabular}[c]{@{}c@{}}CPU stalls caused by\\ memory loads\end{tabular} & 15.16                       & \textbf{40.8}      & 30.15           \\
DRAM Bytes pre Ops                                                          & 0.24                        & \textbf{11.1}      & 0.41           \\ \hline
\end{tabular}
\vspace{-15pt}
\end{table}

\begin{figure}[!t]
\centering
\setlength{\abovecaptionskip}{0pt}
\includegraphics[width=0.85\columnwidth]{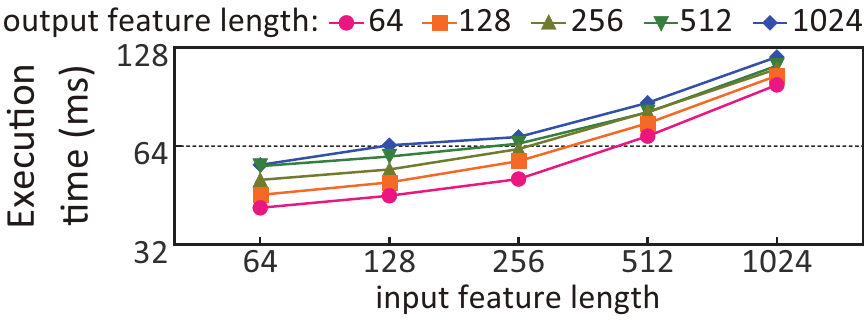}
\caption{Execution time of GCN model on graph with 0.25M vertices and 0.96M edges w.r.t input/output feature length.}
\label{fig:factors}
\vspace{-15pt}
\end{figure}

As GNNs operate on large attributed graphs and the graph structures including input feature dimension, output feature dimension (corresponding to GNN architecture) affects the execution dramatically, we take GCN as an example and further evaluate how these graph features affect the execution time of GNN. Note that a synthetic graph that can be scaled for the evaluation is utilized in the experiment. The experimental result is presented in \autoref{fig:factors}. It reveals that the GNN execution time increases with both larger input feature dimension and output feature dimension while the execution time is more sensitive to the input feature dimension. For instance, when input feature dimension changes from 64 to 1024, the execution time increases by 2.21X. However, it increases by only 1.32X when the output feature goes up from 64 to 1024. Meanwhile, we observe that the graph convolution operation can be symmetric in aggregate with sum operator cases and we may exchange the input feature dimension and output feature dimension. The proof will be detailed in \autoref{sec:optimization}. With these observations, we may improve the computing efficiency by exchanging the input and output features without affecting the GNN processing.

\subsection{Hardware architecture for GNNs}
The state-of-art graph learning frameworks such as DGL, PyG essentially rely on general purposed processors (GPPs) i.e. CPU and GPU. Nevertheless, GPPs especially GPUs fail to take advantage of a large amount of parallel processing engines on GNNs that involve a large amount of irregular traverse and computing over large sparse graphs. As a result, GPUs suffer workload imbalance, memory divergence, and branch divergence for GNN processing \cite{GraphCache, HyGCN, GCN_Char}. The relatively high performance of GNNs on GPUs is mostly attributed to the extremely high-bandwidth memory, which will incur considerable energy consumption. Unlike GPPs, specialized hardware accelerators promise to offer energy-efficient processing for a specific domain of applications. Great successes have been achieved on neural networks and graph processing that are closely relevant to GNNs. Nevertheless, neither neural network accelerators nor graph processing accelerators can process GNNs with combined neural network processing (feature extraction) and graph processing (aggregate and update). In addition, even for the graph processing part, existing graph processing accelerators assume simple graph structure with fixed scalar feature while the graphs utilized in GNNs usually have much more complex attributes and the attributes change across the different GNN layers, which poses more pressure to the on-chip data buffering and memory access optimization.    

Instead of reusing existing hardware architectures for graph convolution network, the authors proposed HyGCN, a specialized accelerator for GCN processing. They take the hybrid computing pattern of GCN \cite{HyGCN} into consideration and have separate processing modules for the neural network processing and graph-like computing respectively. Nevertheless, HyGCN still fails to unleash the potential of the GNN acceleration on a few aspects. 

First of all, HyGCN mainly assumes a moderate graph with less million vertices and billion edges while many realistic GNN applications targeted at social networks and e-commerce can be orders of magnitude larger. Since there is usually limited on-chip buffer for the accelerator, graph partition, and scheduling over the imbalanced graph partitions on top of the large sparse graphs need to be intensively explored to alleviate the memory access bottleneck as illustrated in the prior subsection. 

Secondly, there is a lack of optimization for the irregular memory accesses caused by the large sparse attributed graph, which plays a key role in general GNN processing. For instance, many large graphs extracted from social networks are highly skewed \cite{GraphCache}. The analysis of datasets used by this work indicates the top 20\% vertices with higher degree are connected to the 50-85\% edges of the whole graph. The vertex property of these high-degree vertices are more likely to be reused during the aggregation. In contrast, the low-degree vertices are less probably to be reused. These skewed vertices are equally buffered in HyGCN, which can lead to frequent data movement between the on-chip buffer and the external DRAM. While the feature dimension is usually large in GNNs, this further deteriorates the memory access efficiency of HyGCN. 

Finally, HyGCN has separate the modules for the regular neural network processing part and irregular graph processing part. Accordingly, they need independent on-chip buffers which consume considerable chip area. Although they can be pipelined, the imbalanced computing of the different processing stages as shown in the prior subsection makes it difficult to make use of both modules for general GNN processing efficiently. We argue that a unified hardware design that can reuse the limited on-chip buffer among the different processing stages can provide more energy-efficient GNN processing. 

In summary, GNNs that combine both neural network processing and graph-like processing can be computing bound and memory bound. Particularly, the processing bottleneck changes with the GNN algorithms and targeted graphs, which makes general GNN accelerator design rather challenge. The unique combined computing features also make GNN processing inefficient on GPPs and hinders us to reuse the existing DNN accelerators and graph processing accelerators. The state-of-the-art accelerator HyGCN which focuses on GCN acceleration still fails to consider the influence of the large sparse graphs on GNN processing sufficiently and to unleash the potential of GNN acceleration. This motivates us to concentrate on the memory access optimization for energy-efficient general GNN processing in this work.

\section{EnGN Architecture}\label{sec:EnGN_architecure}
\subsection{EnGN hardware architecture}
\begin{figure}[!t]
\centering
\includegraphics[width=0.99\columnwidth]{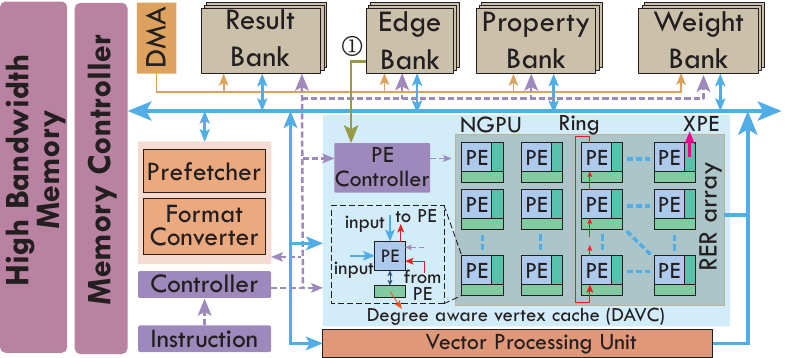}
\caption{EnGN hardware architecture.}
\label{fig:EnGN hardware architecture}
\vspace{-10pt}
\end{figure}
On top of the unified EnGN processing model, we develop a customized EnGN accelerator as shown in \autoref{fig:EnGN hardware architecture}. It adopts 32-bit fixed point to maintain the accuracy of GNN inference and integrates a neural graph processing unit (NGPU) to perform Feature extraction, Aggregate, and Update operation in a unified architecture. It has an array of homogeneous processing elements (PE) and the array size is $128 \times 16$. Each PE unit contains a local register file to store the temporary results and acts as intermediate for inter-PE communication. Each PE in the same column of the Ring-Edge-Reduce (RER) array is connected to its neighbors in a ring network to perform aggregate operation and each PE in the same row of the RER array can process a vertex property, which means the NGPU can process 128 vertices simultaneously. However, such processing parallelism requires substantial memory bandwidth. Thereby, to avoid performance degradation, EnGN optimizes the memory access patterns for vertex data and edge data moving. For source vertex data access in the large graph, we adopt the graph tiling technique and ensure that the source vertex fetching only induces accesses to the continuous memory addresses. For random destination vertex accesses in the aggregate and update stage, EnGN leverages the hashed edge data layout and multi-level cache method to avoid write conflicts and improve data hit rate in the compact on-chip buffer. During processing, the edge parser of NGPU reads the edge list of the graph from the edge banks and parses it into bit-stream that controls the PE-array to perform inter-row aggregate operation (\ding{172} in~\autoref{fig:EnGN hardware architecture}). In addition, as shown in~\autoref{fig:EnGN hardware architecture}, each PE in the NGPU is attached by an XPE to perform activation functions, bias operation, and rounding operation in the update stage. A vector processing unit (VPU) is used to deal with different feature extraction, aggregate and update functions of GNNs illustrated in~\autoref{table:GNN algorithms on GNN framework}. Two auxiliary modules: Prefetcher and Format converter, are used to assist the memory accesses and improve the input graph format compatibility respectively.

\subsubsection{The RER PE array}
\begin{figure}[!t]
\centering
\setlength{\abovecaptionskip}{5pt}
\includegraphics[width=0.85\columnwidth]{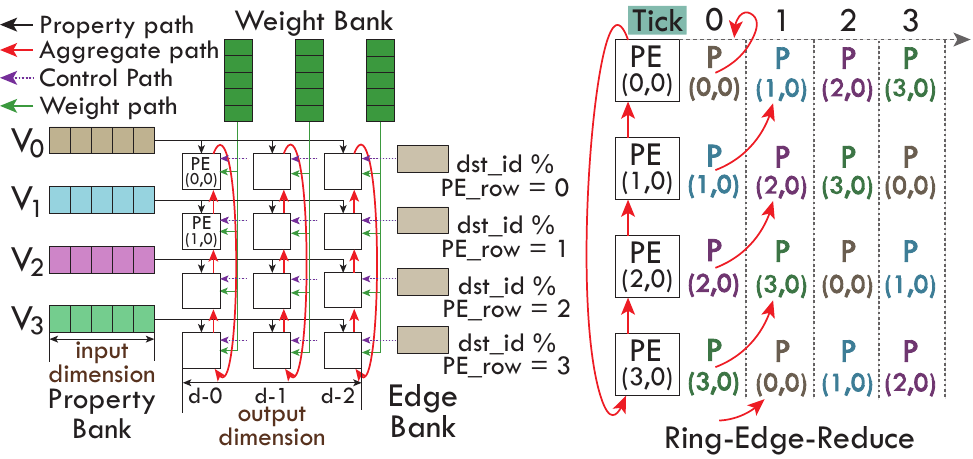}
\caption{Architecture details.}
\label{fig:Architecture details}
\vspace{-20pt}
\end{figure}

The feature extraction stage maps the high dimensions property of vertices to the low dimensions by using the learned weight matrix, and this stage is simply matrix multiplication operation. As shown in~\autoref{fig:Architecture details}, in order to handle the arbitrary-dimension property of GNN algorithms, we propose the graph property aware (GPA) dataflow to decouple the input property of the vertex and the hardware computing structure. In this manner, each PE in the same column of PE-array is responsible for a single dimension of vertex property and each PE in the same row handles a single vertex. The properties of a vertex are arranged in columns and aligned in the property bank. The dimensions of input vertex property become independent to the hardware architecture and can be continuously injected into the PE-array regardless of the array size and the property dimension. In this manner, the processing unit can handle vertex with arbitrary dimension property.

\subsubsection{The RER topology for PE communication} 
The aggregate procedure needs to collect the information according to the edge information. Thereby, as shown in~\autoref{fig:Architecture details}, each row of the PE-array in NGPU possesses a dedicated edge bank and each PE in the same row receives the same control signal parsed from edge list in the graph to gather the corresponding vertex property. Meanwhile, because each PE needs to broadcast its own vertex features generated by the feature extraction stage to all other PEs in the same column, aggregating the received information simultaneously can result in a large amount of hardware resource and power consumption. Thereby, inspired by the ring-all-reduce concept~\cite{ringreduce}, we propose the ring-edge-reduce (RER) aggregate dataflow to conduct aggregate stage inside the PE array instead of moving the data out to the buffer. As shown in~\autoref{fig:Architecture details}, because each column of PE performs the same operations without any communication in between, each PE in the same column of the array is connected to its neighbors through an on-chip network of ring topology. Each PE in the same column only communicates with its two nearest neighbors (north, south). In our design, the PE sends its data to the northern neighbors and receives the data sent from the southern neighbors for property aggregating. In this manner, a PE can select the relevant vertices to aggregate based on the control signal parsed from the edges during the data flow across the ring. 

\begin{figure}[!t]
\centering
\setlength{\abovecaptionskip}{5pt}
\includegraphics[width=0.9\columnwidth]{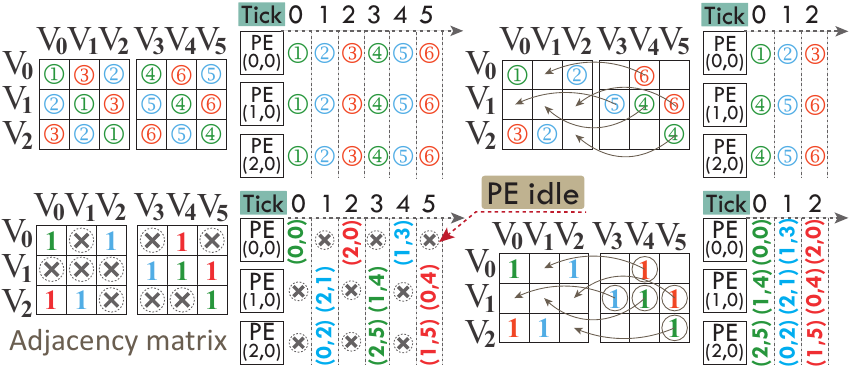}
\caption{Edge reorganization.}
\label{fig:edge_rearrange}
\vspace{-20pt}
\end{figure}

The RER dataflow makes the hardware design simple yet efficient when the graph is dense and the vertex properties that flow through the ring are mostly used for aggregation. However, many of the large graphs in practice are sparse and aggregation in PEs is inactive in many cases. A RER dataflow example on a sparse graph and the adjacency matrix is shown in~\autoref{fig:edge_rearrange}. The computing array is assumed to be $3 \times 3$. In cycle 0, three edges from different edge banks will be fetched and the properties of $V_0$, $V_1$, and $V_2$ will flow across the ring at the same time. It takes the RER three cycles to complete the movement of the three vertex properties and the corresponding aggregation on $V_0$, $V_1$, and $V_2$. Similarly, it takes the RER another three cycles to repeatedly transfer the three vertex properties through the ring to aggregate on $V_3$, $V_4$, and $V_5$. Thereby, it takes the RER at least 6 cycles to perform the aggregate of the graph and many of the time slots are idle as marked with crosses in the figure. 

To improve the efficiency of the aggregation, we further analyze the reason for the idle time slots. For example, PE(1, 0) is idle in Cycle 0 because the edge to be processed is $2\rightarrow1$ and it does not have the properties of vertex 2 yet. However, if it fetches the edge $1\rightarrow4$ first, it can perform the aggregate of vertex 4 using the property of vertex 1 at Cycle 0. With this observation, we propose to reorganize the edges in each edge bank to ensure the vertex properties flowing through the ring is used as much as possible. \autoref{fig:edge_rearrange} exhibits the reorganized edges and the corresponding aggregation. With the edge reorganization, the aggregate completes in 3 cycles and the computing array is fully utilized. Basically, the order of the vertex properties flowing through the ring is known given the computing array. The required vertex property of each edge is also determined. Thereby, reorganizing the edges in each edge bank based on the order of the vertex properties flowing through the ring can maximize the aggregation efficiency of the computing array.  


\subsection{The On-chip Memory Hierarchy}
\begin{figure}[!t]
\centering
\setlength{\abovecaptionskip}{5pt}
\includegraphics[width=0.9\columnwidth]{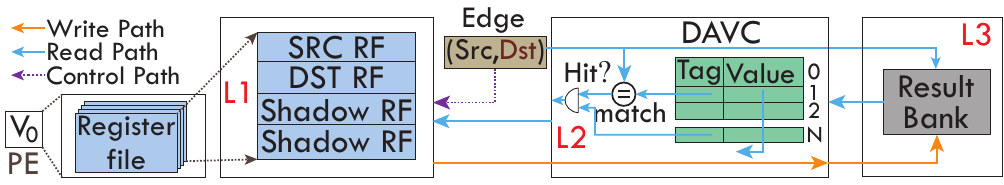}
\caption{Memory hierarchy.}
\label{fig:Memory hierarchy}
\vspace{-20pt}
\end{figure}

\textbf{PE register file} The register files (RF) equipped in the PEs are divided into four groups including source vertex groups(SRC RF), destination vertex groups (DST RF), and two shadow groups (Shadow RF), which is depicted in~\autoref{fig:Memory hierarchy}. The SRC RF stores the source vertex values generated in the feature extraction stage. The DST RF stores the destination vertex feature updated during the aggregate and update stages. In addition, there are two programmer-invisible Shadow RFs holding the SRC and DST vertex values previously generated by the PEs of the same column.

\textbf{Multiple-level caches} The real-world graph has up to millions of vertices. Although the graph tiling technique adopted by EnGN helps fit the sub-graphs into the on-chip buffer, the set of vertices in the sub-graphs will still outsize the register files of the PE array. Meanwhile, the result banks are used to store the temporary aggregate results. PE frequently accesses the long-latency result bank will result in performance degradation. Consequently, as shown in~\autoref{fig:Memory hierarchy}, we insert a degree aware vertex cache (DAVC) between the result banks and the register file of each PE to improve the performance of the EnGN. The register file, DAVC, and the result banks are regarded as the first, second, and last level memories on-chip, respectively. All capacity of DAVC is used to cache high-degree vertices. The reason will be illustrated in~\autoref{sec:evaluation}. DAVC uses the destination vertex id of edges as the line tag to determine whether the access to the vertex data hit or not in the DAVC. If hit, the vertex data will be directly read to DST RF in the PE unit. Otherwise, EnGN will access the last-level result banks. In this manner, the DAVC can alleviate the overhead incurred by the result bank accesses.
\section{ENGN OPTIMIZATION}
\label{sec:optimization}
\subsection{Observations of GNN computing}
To further optimize the EnGN design, we try to explore the characteristics of GNN algorithms and seek key observations 
that may guide the EnGN architecture optimization. 
Suppose the input graph $G = (V, E)$ with $N$ vertices and $E$ edges is 
depicted with an adjacency matrix $A \in \mathbb{R}^{N \times N}$. The vertex property of the graph is $X\in \mathbb{R}^{N \times F}$ 
with $F$ channels and the learned filters i.e. weight is $W \in \mathbb{R}^{F \times H}$ where $H$ is output property dimension. 
Then, the output of the GNN i.e. $O$ can be represented as~\autoref{GCN_ob}:
\begin{equation}\label{GCN_ob}
\setlength{\abovedisplayskip}{3pt}
\setlength{\belowdisplayskip}{3pt}
O = \sigma(A(XW))
\end{equation}


According to the formulation of GNNs, we obtain two major exploitable observations:

\hangafter 1 \hangindent 1em \noindent \textit{1. The order of \textit{feature extraction} processing and \textit{aggregate} processing in GNNs are exchangeable when the operator in \textit{aggregate} processing is \textit{sum}.}

When the operator used in \textit{aggregate} is \textit{sum} which is widely adopted in GNN algorithms, 
the computing in~\autoref{GCN_ob} can be changed to~\autoref{GCN_ob2} without 
affecting the result because of matrix multiplication associative law. While the amount of operations using 
the distinct computing order is also different, we may choose the order that incurs less computation in each iteration.    
\begin{equation}\label{GCN_ob2}
\setlength{\abovedisplayskip}{3pt}
\setlength{\belowdisplayskip}{3pt}
O = \sigma((AX)W)
\end{equation}

\hangafter 1 \hangindent 1em \noindent  \textit{2. The weight size of GNNs is independent to the size of the input graph and it is usually small. While the input graphs can be large and typically dominate the memory accesses.}

According to~\autoref{GCN_ob}, the weight size of GNNs is irrelevant to the number of vertices in the graph. In this case, the weight size can be much smaller compared to the graphs that may include millions of vertices, which is also a key distinction from CNNs. Input graphs will dominate the memory accesses and dealing with the large graphs in GNNs will be critical to the accelerator performance.


\subsection{Dimension-aware stage re-ordering}\label{sec:dimensionaware}
According to Observation 1, the processing order of GNN stages, the feature extraction, aggregate and update stages, will not affect the computing results, 
but it can change the total number of operations in a GNN. We analyze the quantity of 
operations when using different computing order, and aim to find the best way to schedule the stages.
For \textit{feature\_extraction}, the number of operations i.e. multiply-accumulate 
in~\autoref{GCN_ob} and~\autoref{GCN_ob2} are the same and it is equal to $N \times F \times H$. 
Similarly, \textit{update} does not change with the computing order.
Nevertheless, for \textit{aggregate}, the order of GNN computing leads to different 
number of operations i.e. accumulation in \textit{aggregate}. When~\autoref{GCN_ob} is used, 
the number of operations is $E \times F$. When~\autoref{GCN_ob2} is chosen, the amount of operations 
becomes $E \times H$. 

While the property dimension varies as observed in last subsection, $F$ is not equal to $H$. 
To reduce the total computing, when the input vertex property dimension $F$ 
is larger than output feature dimension $H$, we should choose ~\autoref{GCN_ob} for GNN computing. 
Otherwise, we should use~\autoref{GCN_ob2}. Following this idea, we propose a 
dimension-aware stage reordering (DASR) strategy based on the input and output property 
dimension comparison. The DASR can be implemented by altering the 
instruction sequence that defines the computing order of GNNs, so it will not incur 
additional hardware overhead.

\begin{figure}[!t]
\centering
\setlength{\abovecaptionskip}{0pt}
\includegraphics[width=0.9\columnwidth]{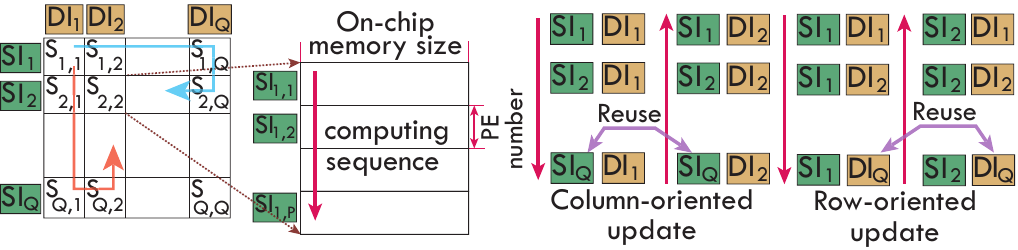}
\caption{Graph tiling and tile scheduling.}
\label{fig:Hierarchy tiling}
\vspace{-20pt}
\end{figure}

\subsection{Graph tiling and scheduling}\label{sec:graphtiling}
According to Observation 2, a real-world graph that can be very large dominates the memory accesses in GNNs and it cannot be fitted to the limited on-chip memory of EnGN. To address this issue, EnGN tiles the large graph into intervals and shards using a grid partition approach proposed in~\cite{GridGraph}. The basic idea of the grid partition is to divide all the vertices into $Q$ disjointed intervals. Then the edges of the graph with both source and destination vertices limited to one interval can be partitioned into $Q^{2}$ disjointed shards. Each shard must be fitted to the on-chip memory of EnGN to ensure efficient computing without external memory accesses. 

With the tiling, EnGN processes with the granularity of a tile. For each tile,
the number of vertices remains larger than the row size of the PE array while 
each row of PE can only handle a single vertex at one time according to the dataflow 
proposed in prior section. In this case, the vertices are processed in batch and 
the batch size is equal to the row size of the PE array. The batch processing of a tile is 
described in~\autoref{fig:Hierarchy tiling}.
Instead of conducting \textit{feature\_extraction} and \textit{aggregate} 
sequentially, we have them overlapped. Basically, \textit{aggregate} starts when a 
batch of vertices complete \textit{feature\_extraction}. 

Although tiling ensures EnGN to process using just the data that are accommodated in the 
on-chip buffers, there are still data dependency between the different tiles. 
The order of the tile execution essentially affects the data reuse and the 
amount of external memory accesses accordingly. Thereby, tile scheduling is also 
an important design option that needs to be intensively optimized.

The graph is split into a 2D array of tiles. The tiles in each row have the same source vertices 
while the tiles in the same column have the same destination vertices. Intuitively, we may schedule in 
either a row manner or a column manner. In the column-major order, new source vertices must be reloaded 
tile by tile while the destination vertices in the same interval reside in on-chip buffer until the 
column of tiles complete execution. 
In the row-major order, source vertex properties can be buffered until the whole row of tiles is processed. We also notice that there are also shared data between neighboring columns or rows and propose to schedule with an S-shape 
as shown in~\autoref{fig:Hierarchy tiling}. For example, the bottom tile of a column shares the same source 
vertices with the bottom tile in the next column. Similar data sharing can be observed in row manner.


The different tile scheduling strategies mainly differ on the external memory accesses and we 
quantitatively analyze the I/O cost. For column-major order, each column of tiles requires to load 
$Q$ tiles of source vertices and the total amount of load is $Q^2$. 
When neighboring column data reuse is considered, the amount of data to be loaded becomes $Q^2-Q+1$.  
While the destination vertices in each column can be reused, the total amount of write is $Q$.
For row-major order, the amount of read is the same, but the amount of write is much larger, because 
tiles in a row generate many intermediate outputs and must be frequently swapped to external 
memory among different tile execution. The total amount of write is $Q^2$.
While the dimension of the vertex property also affects the amount of I/O cost and the dimension of 
input vertex property and output vertex property is usually different, we further take the vertex property dimension into consideration and the I/O cost is summarized 
in~\autoref{table:io_cost}.

\begin{table}[!t]
    \centering
    \footnotesize
    \setlength{\abovecaptionskip}{0pt}
\caption{I/O cost.}
\label{table:io_cost}
\begin{tabular}{ccccc}
\hline
 &Read Size        &Write Size \\\hline
Column-oriented       &$(Q^2-Q+1)F+QH$    &$QH$   \\
Row-oriented          &$QF+(Q^2-Q+1)H$    &$Q^2H$ \\\hline
\end{tabular}
\vspace{-20pt}
\end{table}



Suppose that the latency of read and write external memory is equal. Comparing the overhead of the two different tile 
scheduling strategies, we obtain the following formulation:
\begin{equation}\label{equ:io_cost}
\small
\setlength{\abovedisplayskip}{3pt}
\setlength{\belowdisplayskip}{3pt}
\begin{split}
      IO_{column-major} - IO_{row-major} \approx (Q-1)(2H-F)>&0
\end{split}
\end{equation}

Based on~\autoref{equ:io_cost}, it can be concluded that the column-major order scheduling outperforms 
the row-major order scheduling when F is smaller than 2H. Otherwise, row-major order scheduling is preferred.
While $F$ and $2H$ are mostly determined by the GNNs and the comparison varies, we employ an adaptive scheduling
to minimize the external memory accesses. The adaptive scheduling option is explicitly encoded in the instructions 
which are generated at compilation time based on the GNN models.

\section{EVALUATION}\label{sec:evaluation}

\begin{table*}[!tb]
    \centering
    \scriptsize
    \setlength\tabcolsep{4pt}
    \setlength{\abovecaptionskip}{0pt}
    \caption{System configurations.}
    \label{table:system configurations}
\begin{tabular}{cccccc}
\hline
\textbf{}                  & \textbf{CPU-DGL/PyG}                                        & \textbf{GPU-DGL/PyG}                                           & \textbf{HyGCN}                                                                      & \textbf{EnGN\_22MB}                                                               & \textbf{EnGN}                                                                     \\ \hline
Compute Unit               & \begin{tabular}[c]{@{}c@{}}3.0GHz @\\ 65 cores\end{tabular} & \begin{tabular}[c]{@{}c@{}}1.25GHz @\\ 5120 cores\end{tabular} & \begin{tabular}[c]{@{}c@{}}1GHz @ 32 SIMD 16 cores \\ and 32X128 arrays\end{tabular} & \begin{tabular}[c]{@{}c@{}}1GHz @ 128X16 arrays\\ 32 PE units in VPU\end{tabular} & \begin{tabular}[c]{@{}c@{}}1GHz @ 128X16 arrays\\ 32 PE units in VPU\end{tabular} \\
On-chip Memory             & 42.75MB                                                     & 34MB                                                           & 22MB+128KB                                                                          & 22MB+128KB                                                                        & 1600KB                                                                            \\
Off-chip Memory            & 255.9GB/s DDR4                                              & $\sim$900GB/s HBM 2.0                                          & 256GB/s HBM 1.0                                                                     & 256GB/s HBM 2.0                                                                   & 256GB/s HBM 2.0                                                                   \\
Peak Performance (GOP/s)   & -                                                           & -                                                              & 8704                                                                                & 6144                                                                              & 6144                                                                              \\
Area (mm2)                 & -                                                           & -                                                              & 7.8 (12nm)                                                                          & 31.2 (14nm)                                                                       & \textbf{4.54 (14nm)}                                                              \\
Power (W)                  & 150                                                         & 300                                                            & 6.7                                                                                 & 10.2                                                                              & \textbf{2.56}                                                                     \\
Energy Efficiency (GOPS/W) & -                                                           & -                                                              & 1.30                                                                                & 0.61                                                                              & \textbf{2.40}                                                                     \\
Area Efficiency (GOPS/mm2) & -                                                           & -                                                              & 1.16                                                                                & 0.20                                                                              & \textbf{1.35}                                                                     \\
GNN speedup on average     & -                                                           & -                                                              & 1                                                                                   & 5.44X                                                                             & \textbf{2.97X}                                                                    \\ \hline
\end{tabular}
\vspace{-10pt}
\end{table*}

\begin{table}[!tb]
    \centering
    \scriptsize
    \setlength\tabcolsep{4pt}
    \setlength{\abovecaptionskip}{0pt}
    \caption{GNN models and Datasets.}
    \label{table:dataset}
\begin{tabular}{c|ccccc}
\hline
Model                                           & Graph          & \#Vertices                & \#Edges                    & \begin{tabular}[c]{@{}c@{}}\#Feature/\\ \#Relation\end{tabular} & Label                   \\ \hline
\multicolumn{1}{c|}{\multirow{3}{*}{GCN}}       & Cora (CA)~\cite{Cora}       & 2708                      & 10556                      & 1433                                                             & 7                       \\
\multicolumn{1}{c|}{}                           & PubMed (PB)~\cite{Cora}     & 19717                     & 88651                      & 500                                                              & 3                       \\
\multicolumn{1}{c|}{}                           & Nell (NE)~\cite{Nell}      & \multicolumn{1}{c}{65755} & \multicolumn{1}{c}{251550} & \multicolumn{1}{c}{5415}                                         & \multicolumn{1}{c}{210} \\
\multicolumn{1}{c|}{\multirow{3}{*}{GS-Pool}}   & CoraFull (CF)~\cite{Cora-full}   & 19793                     & 126842                      & 8710                                                             & 67                      \\
\multicolumn{1}{c|}{}                           & Reddit (RD)~\cite{Reddit}     & 232965                    & 114.6M                     & 602                                                              & 41                      \\
\multicolumn{1}{c|}{}                           & Enwiki (EN)~\cite{NeuGraph:Parallel}      & 3.6M                      & 276.0M                     & 300                                                              & 12                      \\
\multicolumn{1}{c|}{\multirow{3}{*}{Gated-GCN}} & Amazon (AN)~\cite{NeuGraph:Parallel}     & 8.6M                      & 231.6M                     & 96                                                               & 22                      \\
\multicolumn{1}{c|}{}                           & Synthetic A (SA)~\cite{RMAT} & 4.19M                     & 67.1M                      & 100                                                              & 16                      \\
\multicolumn{1}{c|}{}                           & Synthetic B (SB)~\cite{RMAT} & 8.38M                     & 134.2M                     & 100                                                              & 16                      \\
\multicolumn{1}{c|}{\multirow{2}{*}{GRN}}       & Synthetic C (SC)~\cite{RMAT} & 12.41M                    & 205.3M                     & 64                                                               & 16                      \\
\multicolumn{1}{c|}{}                           & Synthetic D (SD)~\cite{RMAT} & 16.76M                    & 268.4M                     & 50                                                               & 16                      \\ \hline
\multicolumn{1}{c|}{\multirow{4}{*}{R-GCN}}     & AIFB (AF)~\cite{RGCN}      & 8285                      & 29043                      & 91                                                               & 4                       \\
\multicolumn{1}{c|}{}                           & MUTAG (MG)~\cite{RGCN}      & 23644                     & 192098                     & 47                                                               & 2                       \\
\multicolumn{1}{c|}{}                           & BGS (BG)~\cite{RGCN}        & 333845                    & 2166243                    & 207                                                              & 2                       \\
\multicolumn{1}{c|}{}                           & AM (AM)~\cite{RGCN}         & 1666764                   & 13643406                   & 267                                                              & 11                      \\ \hline
\end{tabular}
\vspace{-15pt}
\end{table}

\subsection{Experimental setup}
\noindent\underline{\textbf{Accelerator simulator}} We built a cycle-accurate simulator to measure the performance of EnGN accelerator. This simulator models each module of EnGN accelerator faithfully and the timing behaviors of the modules are co-verified with the synthesized RTL design. The simulator is also integrated with Ramulator \cite{Ramulator} that supports High Bandwidth Memory (HBM 2.0) to characterize the memory accesses to HBM 2.0 with 256GB/s bandwidth.

\noindent\underline{\textbf{EnGN configuration\&implementation}} The configuration of EnGN is depicted in \autoref{table:system configurations}. We synthesized the EnGN using Synopsys Design Compiler (DC) with the TSMC 14nm process technology, conducted the placing-and-routing using Synopsys ICC compiler (ICC), and estimated the power consumption using Synopsys PrimeTime (PT). The energy of HBM 2.0 is estimated with 3.9 pJ/bit as in\cite{HBM20}.

\noindent\underline{\textbf{Baselines}} We compared the performance and energy efficiency of EnGN with that of three different baseline architectures. The first two are general-purpose processors i.e. CPU and GPU, and the third one is a state-of-the-art GCN accelerator called HyGCN. \underline{\textbf{CPU}} platform is equipped with Intel Xeon(Skylake) 6151@3.0GHz processor and 696GB DRAM and \underline{\textbf{GPU}} platform is equipped with NVIDIA Tesla V100 SXM2 and 32GB HBM2. To make good use of the general-purposed processors, we adopted the state-of-art frameworks i.e. DGL and Pytorch geometric (PyG) to execute the GNN algorithms. The implementations are denoted as CPU-DGL, CPU-PyG, GPU-DGL, and GPU-PyG respectively. \underline{\textbf{HyGCN}} that leverages 22MB eDRAM and specialized computing arrays for GNN processing achieve remarkable performance speedup over the GPU implementations. 
To make a fair comparison with HyGCN, we have EnGN configured with the same amount of on-chip buffer. Due to the lack of 14nm eDRAM library, we replace the eDRAM with SRAM in the experiments. More detailed configurations can be found in \autoref{table:system configurations}.

\noindent\underline{\textbf{GNN models and datasets}} To benchmark the performance of EnGN accelerator, we implemented a set of typical GNN models on two distinct groups of datasets as shown in~\autoref{table:dataset}. The top part includes four algorithms i.e. GCN~\cite{GCN}, GraphSage-Pool (GS-Pool)~\cite{Reddit}, Gated-GCN~\cite{Gated_Conv}, and GRN~\cite{GRN}, which are mainly used for semi-supervised classification. The four algorithms perform on seven real-world graph datasets and four synthetic graph datasets. The bottom part mainly targets at knowledge graph application and R-GCN~\cite{RGCN} is a widely adopted entity classification algorithm. The corresponding datasets are from four typical knowledge graphs. Particularly, note that the feature and label columns represent the dimension of a vertex and the number of labeled classes respectively.

\begin{figure*}[!t]
\centering
\setlength{\abovecaptionskip}{1pt}
\includegraphics[width=1.9\columnwidth]{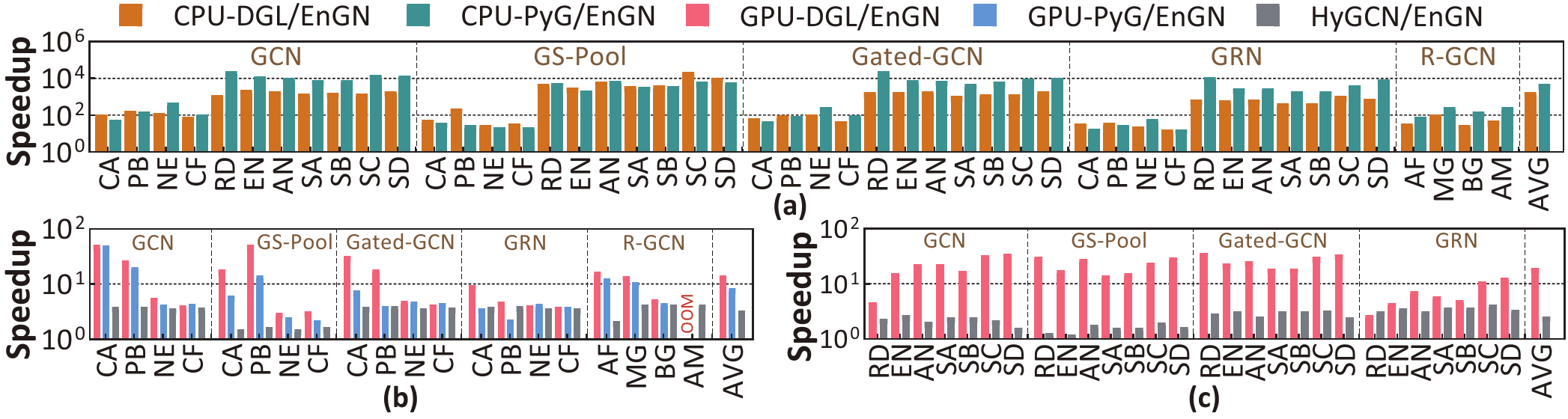}
\caption{Performance comparison of EnGN over CPU, GPU, and HyGCN. (a) Performance speedup of EnGN over CPU-DGL and CPU-PyG. (b) Performance speedup of EnGN over GPU-DGL, GPU-PyG, and HyGCN on small datasets. (c) Performance speedup of EnGN over GPU-DGL and HyGCN on large datasets. Since GPU-PyG runs out of memory (OOM), it is omitted.}
\label{fig:CPU speedup comparison}
\vspace{-10pt}
\end{figure*}

\noindent\underline{\textbf{Evaluation Metrics}} In this experiment, we take the end-to-end inference time of GNNs as the performance metric, billion operations per second (GOP/s) as the throughput metric and billion operations per second per Watt (GOPS/W) as the energy-efficiency metric.
\subsection{Experimental results}
\textbf{Power\&Area} \autoref{table:system configurations} shows the power and area of HyGCN, EnGN\_22MB, and EnGN. As the area of eDRAM is much smaller than SRAMs, the power and area of EnGN\_22MB are larger than HyGCN, but the performance speedup is more than 5X higher. Accordingly, the energy efficiency of EnGN\_22MB is relatively lower in general. Nevertheless, when we compare HyGCN and ENGN, we notice that EnGN still achieves around 3X performance speedup despite the much smaller on-chip buffer. It indicates that the architecture of EnGN greatly lowers the on-chip memory requirements and power consumption. In this case, the overall energy efficiency of EnGN is 1.85X higher. 



\begin{figure*}[!t]
\centering
\setlength{\abovecaptionskip}{1pt}
\includegraphics[width=1.9\columnwidth]{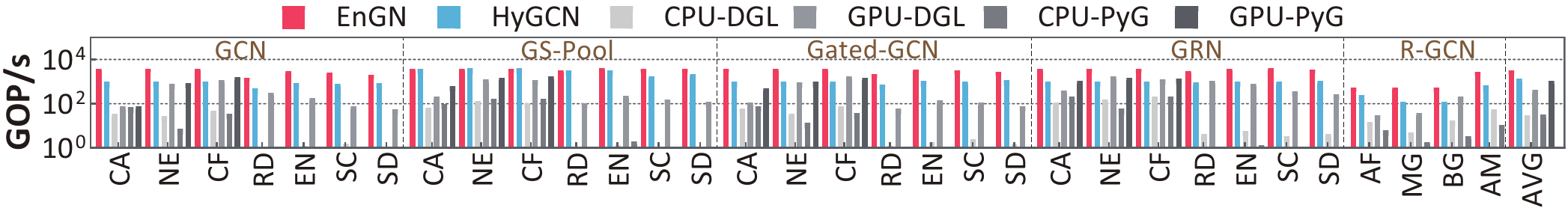}
\caption{Throughput of EnGN, CPU, GPU, and HyGCN. Some datasets are ignored due to literature space constraints.}
\label{fig:throughput}
\vspace{-10pt}
\end{figure*}

\begin{figure*}[!t]
\centering
\setlength{\abovecaptionskip}{1pt}
\includegraphics[width=1.9\columnwidth]{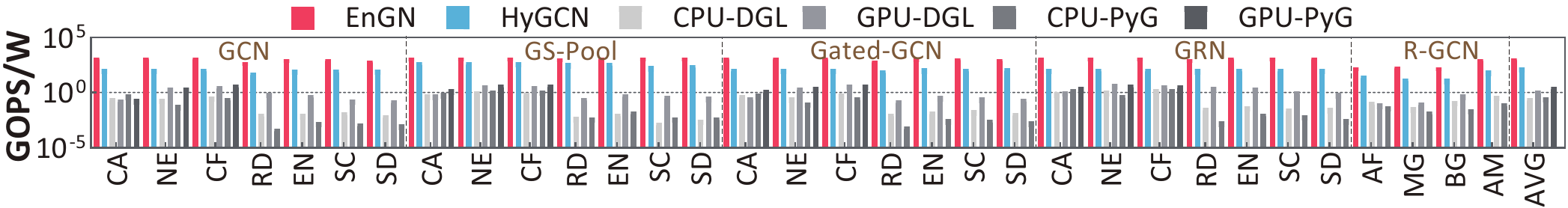}
\caption{Energy efficiency of EnGN, CPU, GPU, and HyGCN. Some datasets are ignored due to literature space constraints.}
\label{fig:power comparison}
\vspace{-20pt}
\end{figure*}

\textbf{Performance} We compare the performance of EnGN to that obtained from the baseline computing platforms including CPU-DGL, GPU-DGL, CPU-PyG, GPU-PyG, and HyGCN. The comparison result is shown in~\autoref{fig:CPU speedup comparison}. The average performance speedup of all the models on all the datasets over CPU-DGL and CPU-PyG are 1802.9X and 5108.4X respectively as shown in the last bar of~\autoref{fig:CPU speedup comparison} (a) denoted as AVG. Also it can be observed that EnGN outperforms CPU in all cases despite the software frameworks, datasets and GNN models. We also compare EnGN with GPU using DGL and PyG respectively. However, PyG runs out of memory on larger datasets due to the lack of sufficient memory optimizations. Thus, we only compare GPU-DGL on large graph datasets as shown in ~\autoref{fig:CPU speedup comparison} (c). On small graph datasets, we have both GPU-DGL and GPU-PyG compared and the comparison is presented in~\autoref{fig:CPU speedup comparison} (b). EnGN gains 14.41X, 8.35X, and 3.33X performance speedup over the GPU-DGL, GPU-PyG, and HyGCN respectively on the small datasets. On large datasets, EnGN achieves 19.75X and 2.61X speedup on average compared to GPU-DGL and HyGCN, respectively. In general, although GPU performs much better than CPU, EnGN still outperforms in all cases.

On top of the computing platforms, we further compare the performance speedup of EnGN on different datasets, it can be noticed that EnGN typically shows significantly higher performance speedup when the dimension of the graph feature is small. For instance, the performance speedup of GS-Pool on SD with smaller feature dimensions is around 10613.17X on CPU-DGL and 35.34X on GPU-DGL while the performance speedup of GS-Pool on CF with the larger feature dimension is less than 36.47X on CPU-DGL and less 2.22X on GPU-DGL. While EnGN with fine-grained dataflow can make good use of the computing resources, the computing efficiency does not vary much with the datasets, which will be illustrated in the following experiments. In contrast, CPUs and GPUs prefer datasets with high-dimension features that can be accessed sequentially and efficiently. Thereby, the different graph features of the datasets lead to distinct performance speedup. Meanwhile, we also find that the performance speedup of EnGN on RD with the relatively high-dimension feature is actually clearly higher than the average performance speedup. The reason for this exception is that RD has rather high average degree than the other graphs. The high-degree graph requires a large memory footprint during the aggregate stage and can no longer be fitted to the on-chip memory or cache. Thereby, the computing efficiency degrades. 

Compared to HyGCN, EnGN achieves higher performance speedup on small datasets than big datasets, because small datasets typically have less on-chip buffer requirements and pose less pressure on the much smaller on-chip buffer of EnGN. In addition, \autoref{fig:CPU speedup comparison} (c) shows the performance speedup of the GS-Pool model on EnGN is lower than other GNN models. This is due to the feature extraction stage of GS-Pool involves more intensive matrix multiplication operations compared to other GNN models and occupies 97\% of the total execution time. Meanwhile, the large systolic array (32X128) of HyGCN performs better than the EnGN with a relatively smaller computing array (128X16) on the feature extraction stage, which conceals the performance improvement brought by the memory hierarchy and scheduling optimization adopted by EnGN.

\textbf{Throughput}~\autoref{fig:throughput} shows the measured throughput of EnGN, CPU, GPU, and HyGCN on the GNN benchmark in~\autoref{table:dataset}. The average throughput of EnGN is 3265.87 GOP/s which achieves 79.7\% of the peak throughput i.e. 4096GOP/s. In contrast, the measured average throughput of CPU-DGL and CPU-PyG is only 29.29 GOP/s and 31.95 GOP/s respectively, which is 111.50X and 102.21X lower. GPU with massive parallel processing units performs much better. The average throughput using GPU-DGL and GPU-PyG is 426.30 GOP/s and 1056.91 GOP/s respectively. Still, the throughput of EnGN is 7.66X and 3.09X higher. This is because GPUs are inherently optimized for compute-intensive workloads with regular execution patterns such as neural networks, but handling the aggregate stage of the EnGN processing model with irregular memory accesses suffers from low efﬁciency. While specialized GNN accelerators achieve much higher throughput than the general-purpose processors, architecture optimization of the accelerators especially the on-chip memory hierarchy optimizations proposed in EnGN can further improve the throughput by 2.34X over HyGCN on average. To gain insight into the computing efficiency on different GNN models and datasets, we measure the computing efficiency of the different computing architectures including EnGN, CPU, GPU, and HyGCN. As shown in \autoref{fig:throughput}, the computing efficiency of EnGN typically keeps steady and does not vary much with the models and datasets while CPU, GPU, and HyGCN are more sensitive and the computing efficiency usually fluctuates with the feature dimension of the graphs as pointed out in prior section.

\begin{figure*}[!t]
\centering
\setlength{\abovecaptionskip}{1pt}
\includegraphics[width=1.9\columnwidth]{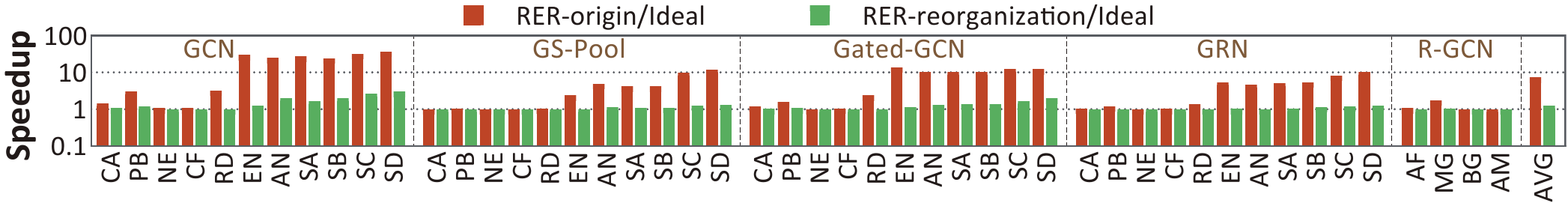}
\caption{Performance comparison of GNNs with original edge layout and reorganized edge layout. Note that both the performance is normalized to the ideal performance with optimal computing resource utilization.}
\label{fig:evaluation_edge_rearrange}
\vspace{-20pt}
\end{figure*}

\textbf{Energy Efficiency} To obtain the energy efficiency of the different computing architectures, we need to measure its power first. The power consumption of CPU and GPU is obtained from the power meter and NVPROF respectively. The power consumption of EnGN is estimated using PrimeTime. The power consumption of CPU, GPU, HyGCN, and EnGN is 150W, 300W, 6.7W, and 2.56W respectively. On top of the power consumption, we further calculated the energy efficiency using the total amount of operations and the execution time. The energy efficiency is shown in~\autoref{fig:power comparison}. The average energy efficiency of EnGN is 1326.35X and 1196.04X higher than CPU-DGL and CPU-PyG respectively. When compared to GPU, the energy efficiency of EnGN over GPU-DGL and GPU-PyG is 213.61X and 133.17X higher on small datasets. The speedup goes up to 529.13X for large datasets on which only DGL can be applied. Meanwhile, the energy efficiency of EnGN is 6.2X higher than HyGCN on average. The great energy efficiency speedup is mainly attributed to the much lower power consumption of the customized EnGN accelerator over the power-hungry general purposed processors and the much higher performance reported in the performance paragraph. The reasons for the higher performance and lower power consumption are already discussed, and we will not dwell on it.

\begin{figure}[!t]
\centering
\setlength{\abovecaptionskip}{1pt}
\includegraphics[width=0.9\columnwidth]{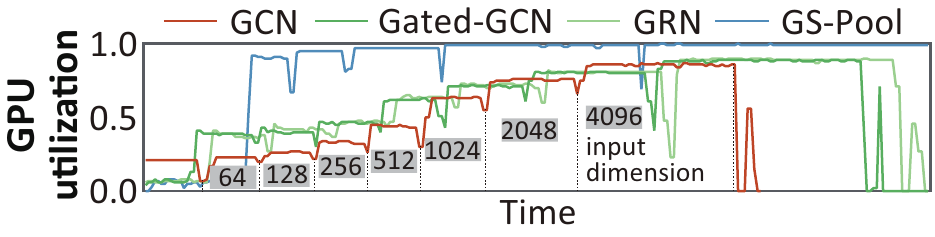}
\caption{GPU utilization w.r.t feature dimensions.}
\label{fig:feature dimension}
\vspace{-20pt}
\end{figure}
\subsection{EnGN optimization evaluation}
\textbf{Edge reorganization and RER} In order to avoid the PE idling in RER, we propose to reorganize the edge list to improve the utilization of the computing array in EnGN. \autoref{fig:evaluation_edge_rearrange} exhibits the performance comparison of GNNs with reorganized edges and original edges. It can be noted that the edge reorganization approach improves the performance significantly and the average performance speedup is 5.4X. Meanwhile, we find that the proposed edge reorganization approach typically works much better for large datasets. The variation of the benefits is mainly caused by the different proportions of aggregation in the total amount of GNN computing. While the aggregation in GNNs dominates the computing when the graph is large, thus the performance improvement is higher. In addition, we have the performance normalized to that of an ideal design which utilizes a fully connected PE column. With the fully-connected topology, the aggregation can achieve optimal performance despite the edge organization. When compared to the ideal topology, the proposed RER topology in combination with the edge reorganization approach achieves near-optimal performance in various cases. In contrast, the hardware design overhead is much smaller compared to the fully connected topology. 

\textbf{Sensitivity to the variation of vertex dimension}
The vertex property dimension varies dramatically in GNNs, so to be insensitive to the vertex property dimension variation is of vital importance to a general GNN accelerator design. In this experiment, we generated a synthetic graph with 65000 vertices, 2.5M edges, and 16 classes. Then we change the input vertex dimension from 64 to 4096 gradually to evaluate the computing efficiency variation under the different vertex property dimension setups. We compare the computing variation of EnGN and GPU-DGL.~\autoref{fig:feature dimension} depicts GPU utilization is lower than 50\% when the vertex property dimension is smaller than 512. Moreover, it drops considerably when GPU threads are wasted under some odd vertex dimension setups. In contrast, the PE utilization of EnGN is irrelevant to the input vertex property dimension because the dataflow in EnGN decouples the input vertex property dimension and the computing array.

\begin{figure}[!t]
\centering
\setlength{\abovecaptionskip}{1pt}
\includegraphics[width=0.9\columnwidth]{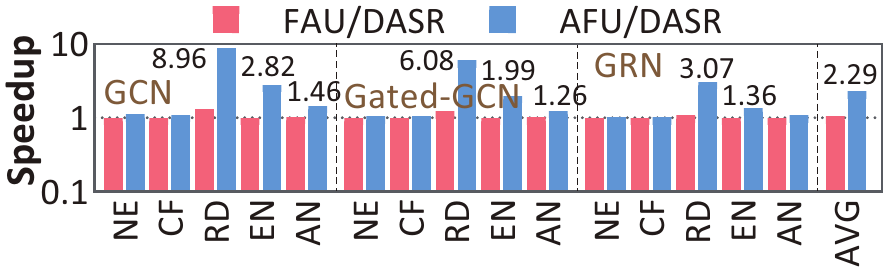}
\caption{Speedup of DASR over FAU and AFU.}
\label{fig:Operatrions comparison}
\vspace{-10pt}
\end{figure}

\begin{figure}[!t]
\centering
\setlength{\abovecaptionskip}{1pt}
\includegraphics[width=0.9\columnwidth]{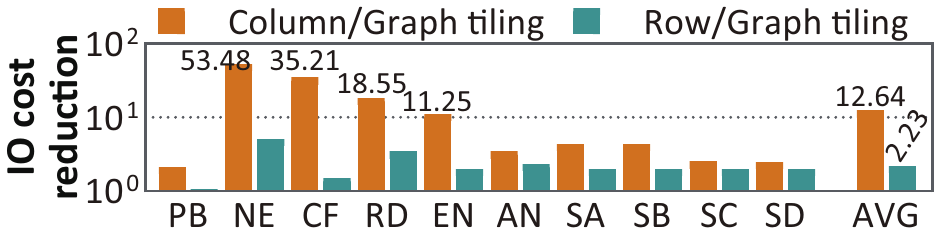}
\caption{I/O cost reduction.}
\label{fig:graphtiling}
\vspace{-20pt}
\end{figure}

\textbf{Dimension aware stage re-ordering} As mentioned in ~\autoref{sec:optimization}, the proposed dimension aware stage reordering technique can reduce the total computing cost. In this evaluation, we get rid of the GS-Pool model because its aggregate stage adopts the average operator which hinders the stage reordering. We compared the performance speedup of EnGN that adopts \textbf{d}imension-\textbf{a}ware \textbf{s}tage \textbf{r}e-ordering (DASR) strategy to two fixed processing strategy: (1) feature\_extraction, aggregate, and update (FAU), and (2) aggregate, feature\_extraction, and update (AFU).~\autoref{fig:Operatrions comparison} illustrates that the DASR strategy can improve the performance of EnGN by 1.047x and 2.297x on averages compared to FAU and AFU, respectively. The reason for the poor performance improvement compared to FAU is the output dimensions of GNN models on most datasets are decreasing, which makes no scheduling necessary. However, in Reddit datasets, our DASR strategy can improve the performance of EnGN by 1.34x and 8.96x compared to FAU and AFU strategy. This is because the output dimensions of vertex property on the last layer are 210 (\autoref{table:dataset}), which is higher than that of on the first layer. When the feature extraction stage performs after the aggregate stage, higher dimensions incurs massive accumulate operators in the aggregate stage. In contrast, when we perform the feature extraction stage before the aggregate stage, the dimension will be compressed to 16 and accumulates operators is only 16 for a vertex in the aggregate stage.

\begin{figure*}[!t]
\centering
\setlength{\abovecaptionskip}{1pt}
\includegraphics[width=1.9\columnwidth]{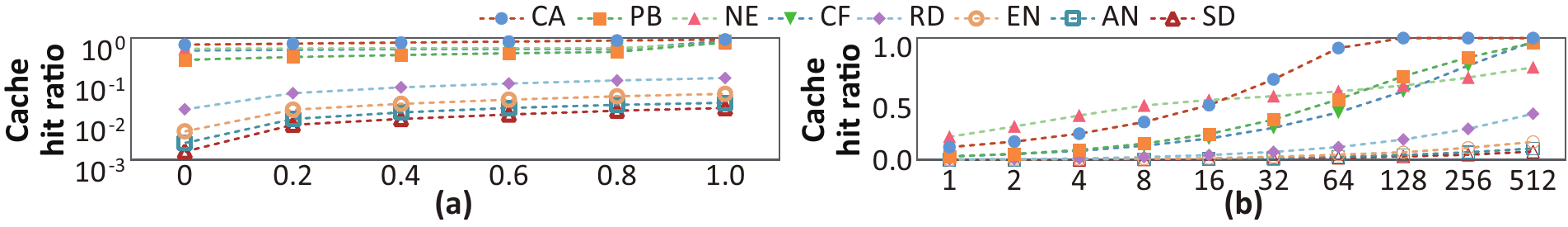}
\caption{Cache hit ratio over different proportions (a) and cache size (KB)(b).}
\label{fig:davc}
\vspace{-10pt}
\end{figure*}

\begin{figure*}[!t]
\centering
\setlength{\abovecaptionskip}{1pt}
\includegraphics[width=1.9\columnwidth]{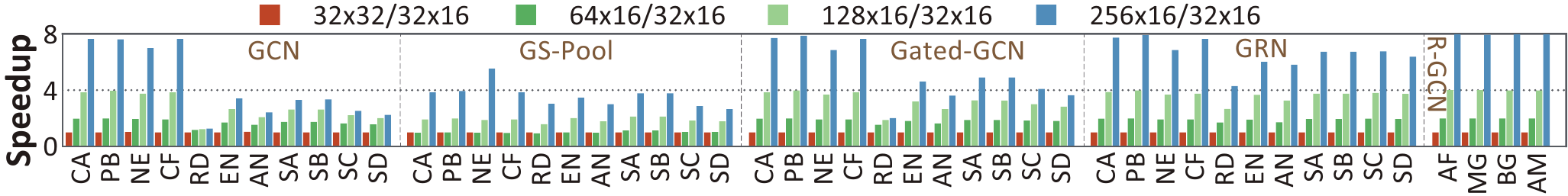}
\caption{Performance over number of PEs.}
\label{fig:PEs}
\vspace{-20pt}
\end{figure*}

\textbf{Graph tiling scheduling} In this evaluation, we leveraged the column-major (Column) and row-major (Row) update strategy as baselines to evaluate our scheduling strategy on GCN model.~\autoref{fig:graphtiling} illustrates the total I/O cost reduction induced by the EnGN scheduling strategy compared to the Column and Row strategies, respectively. In PubMed and large datasets, our graph tiling scheduling strategy only reduces total I/O cost by 3.26x and 1.90x compared to the Column strategy. This is because PubMed and the large dataset only contain $3\sim16$ class labels, which is less than the output dimension of the first layer. In contrast, Nell, Cora-full, and Reddit contain 210, 67, and 41 classes respectively. Thereby, in this case, graph tiling scheduling can reduce the total memory access cost by 29.62x and 3.02x on average when compared to Column and Row, respectively. This is because the Column and Row strategy stick to the fixed policy to update the graph while our graph tiling scheduling can adjust the update dataflow from the Row to Column based on the dimension changes in GNNs.

\textbf{Degree Aware Vertex Cache (DAVC)} DAVC is a standard cache supporting replacement policy like LRU in general. To improve the cache hit rate, we take the vertex degree information into consideration and reserve part of the cache entries for high-degree vertices which are determined with offline static analysis and will not be replaced during the execution. To determine the proportion of the reserved cache entries, we analyze the cache hit rate under various proportion setups ranging from 0 to 1. The experiment in~\autoref{fig:davc} (a) reveals that the cache hit rate increases monotonically with the proportion especially for the larger graphs. The main reason is that on-chip cache is too small relative to the large graphs and thus suffers frequent replacement when LRU policy is applied. Thereby, we have all the cache used for high-degree vertices. Meanwhile, we also analyze the influence of cache size on the cache hit rate. Similar conclusion can be drawn as shown in~\autoref{fig:davc} (b). Basically, the cache hit rate for large graphs remains rather low and larger cache size is preferred. Thus, in order to reduce hardware complexity, the size of DAVC is configured to 64KB.

\subsection{Scalability Analysis}
\textbf{Performance over number of PEs} Since each row of PE array handles one vertex and each column is in charge of one dimension of output property, as the input graph and output property dimensions get larger, the system can be scaled up by adjusting the size of PE-array.
We varied the size of PE-array in EnGN, where the EnGN with $32\times 16$ PE-array is set as baseline.~\autoref{fig:PEs} show EnGN achieves good scalability on all GNN models and datasets. With the increase of the row number in PE-array, the throughput of EnGN is increasing. However, $32\times 32$ array exhibits no improvement over the baseline. This is because the output property dimensions of the first layer (16) on all models are below the column number of PE array (32), which causes underutilization of PE array. Thereby, we can adjust the size of PE array according to the datasets and the complexity of GNN models to maximize the throughput of EnGN.~\autoref{fig:PEs} also witnessed the performance speedup on large datasets is lower than on small datasets. This is due to the large data has higher edge-to-vertex ratio compared to small datasets, which makes the aggregated stage new bottleneck in large PE arrays.
\section{RELATED WORK}
\subsection{GNNs software framework}
There is a large amount of work that aims at building an efficient system for graph applications on single node-machines (CPUs)~\cite{Pregel}, distributed systems~\cite{NUMA-aware}, and GPUs~\cite{PowerGraph}. However, these graph processing frameworks aim at traditional algorithms, and there is a lack of support for graph neural network computation. Even though TuX2~\cite{Gemini} aims to bridge the gap between graph and traditional machine learning algorithms, it is still unable to support the inference and training stage of emerging GNN algorithms. Thereby, NeuGraph~\cite{NeuGraph:Parallel} is proposed to recast the graph specific optimization as dataflow optimization based on Tensorflow. Meanwhile, ~\cite{GraM} published a geometric learning library for deep learning on irregularly structured input data based on Pytorch. The deep graph library~\cite{DGL} provides a fast implementation of GNN models based on PyTorch and MxNet. NeuGraph, Pytoch geometric, and DGL are generally running on the power-hungry CPU and GPUs, which incurs energy-efficient issues and high cost. More importantly, GPUs suffer from the under-utility of stream processors during parallel GNN computation because of the impact of the irregular graph data structure, which makes energy-efficient issues more serious. Thereby, to address these issues, we build an EnGN accelerator designed for large GNNs to support energy-efficient GNN processing.

\subsection{Deep learning \& Graph accelerator}
The resurgence of deep neural network (DNN) and its substantial progress in various applications including image, video, and speech spur the flourishing of the DNN hardware accelerator~\cite{DNN_survey}. For example, Diannao ~\cite{Diannao} maps DNN onto an array of multiply-add units and employs a data tiling policy to exploiting the locality in the parameters. EIE~\cite{EIE:} performs inference using compressed technique and accelerates the inherent modified sparse matrix-vector multiplication. However, these DNN accelerators are designed for traditional DNN such as CNN and RNN, which cannot handle GNNs because they lack the graph propagation model on the accelerator.

The wide gap between the general-purpose architectures and the unique features of graph processing promotes the rapid development of graph processing-specific accelerators based on FPGA and ASIC. For example, Graphicionado~\cite{Graphicionado} and~\cite{Micro} presented a domain-specific accelerator for graph analytics based on a well-defined, popular vertex programming model. However, traditional graph accelerators are designed for traditional graph algorithms, it lacks the computation abstraction required by the neural network, such as tensor and activation operations. Thereby, HyGCN \cite{HyGCN} abstracted the execution flow of GCN into aggregation and combination stage and leveraged the SIMD and systolic arrays to support neural network computation and graph propagation model simultaneously.

\section{Conclusions}
In this paper, we present a high-throughput and energy-efficient accelerator EnGN specialized for large graph neural network processing. In order to provide high throughput processing ability and solve the arbitrary dimension change issues in the GNN algorithms, we proposed ring-edge-reduce update dataflow and the accompanied hardware architecture of RER PE-arrays is designed to simultaneously conduct high-throughput processing in the feature-extraction, aggregate and update stages on GNNs. Meanwhile, the proposed graph tiling and scheduling technique cooperating with a well-designed three-level memory hierarchy enable EnGN to process large graphs efficiently. Experimental results show that EnGN achieves 2.97X speedup while consuming 6.2X less energy compared to the state-of-the-art GCN accelerator HyGCN. Meanwhile, EnGN achieves performance gains of 1802.9X and 19.75X and energy efficiency of 1326.35X and 304.43X compared to CPUs and GPUs on average, respectively.

\ifCLASSOPTIONcompsoc
  \section*{Acknowledgments}
\else
  \section*{Acknowledgment}
\fi
This work was supported in part by the National Natural
Science Foundation of China under Grant 61874124, Grant
61876173, Grant 61432017, Grant 61532017, Grant 61772300, and YESS hip program No.YESS2016qnrc001. Corresponding authors are Ying Wang and Huawei Li.

\ifCLASSOPTIONcaptionsoff
  \newpage
\fi



%
\bibliographystyle{IEEEtran}
\bibliography{IEEEabrv,ref}



%







\end{document}